\documentclass[11pt,a4paper]{article}
\pdfoutput=1
\usepackage{jheppub}
\usepackage[utf8]{inputenc}
\usepackage{amssymb, amsmath}
\usepackage[normalem]{ulem}
\usepackage{hyperref}

\newcommand{\bea}{\begin{eqnarray}}
\newcommand{\eea}{\end{eqnarray}}

\renewcommand{\(}{\left(}
\renewcommand{\)}{\right)}

\newcommand{\Mpl}{M_{\textrm{Pl}}}
\def\H{\mathrm{H}}
\def\V{\mathrm{V}}
\def\e{\mathrm{e}}
\def\be{\begin{equation}}
\def\ee{\end{equation}}

\begin{document}
		
\title{Avoiding the string swampland in single-field inflation: Excited initial states} 

\author{Suddhasattwa Brahma$^{a,*}$ and}
\emailAdd{suddhasattwa.brahma@gmail.com$^*$}

\author{Md. Wali Hossain$^{a,\dagger}$}
\emailAdd{wali.hossain@apctp.org$^\dagger$}
\affiliation{$^{a}$Asia Pacific Center for Theoretical Physics,	Pohang 37673, Korea}

\date{September 22, 2018}

\abstract{One class of single-field inflationary models compatible with the recently-conjectured Swampland criteria would be those in which a Hubble slow-roll arameter $\epsilon_\H$ is not the same as $\epsilon_\V \sim (V'/V)^2$. However, a roadblock for these models (with a convex potential) lie in the unacceptably high tensor-to-scalar ratio, $r$, generically predicted by them. In this work, illustrating through an explicit example, we point out that having a non-Bunch-Davies component to the initial state of cosmological perturbations makes the value of $r$ compatible with observations. In this way, we lay down a new path even for standard models of slow-roll inflation to be consistent with the Swampland criteria by invoking deviations from the Bunch-Davies initial state.}

\maketitle

\section{Motivation} 

Inflationary model-building, thus far, has remained largely unconstrained by theoretical considerations. Recently, it has been pointed out \cite{Vafa:2005ui} that string theory imposes strong restrictions on the type of ($4$-dimensional)  effective field theories allowed as a consistent starting point. One not only has a vast string landscape of vacua of consistent low-energy effective theories but also a much larger swampland enclosing this landscape, the latter being a space of field theories not compatible with quantum gravity. This is a very encouraging development, especially for cosmology, since it implies that `not-anything-goes' as far as inflationary model-building is concerned. In order to be in the landscape, and not in the swampland, an inflationary model needs to satisfy additional theoretical consistency conditions such as the Weak Gravity Conjecture \cite{ArkaniHamed:2006dz}. 

The cosmological implications for having such a swampland are rather remarkable such as the absence of stable de-Sitter (dS) vacua in critical String theory \cite{Obied:2018sgi,Andriot:2018wzk}. In particular for early-universe cosmology, using the criteria for not being in the swampland, one can put severe constraints on the plethora of inflationary models available on the market \cite{Agrawal:2018own}. Indeed, there has been some claims that these criteria, along with observational constraints, rule out all single-field models of inflation \cite{Singlefield1, Singlefield2, Garg:2018reu}. Here, we show that this is not necessarily the case with an explicit mechanism involving excited initial states. 

For completeness, we begin by restating two Swampland criteria (in reduced Planck units):
\begin{itemize}
	\item S1: The range of excursion of a scalar field is bounded from above by $|\Delta \varphi| < \Delta \sim \mathcal{O}(1)$.
	\item S2: The slope of the scalar field potential satisfies a lower bound $|V'|/V > c \sim \mathcal{O}(1)$ when $V>0$. 
\end{itemize}
The first conjecture (S1) states that when one traverses a canonical distance\footnote{The metric for the kinetic term provides a notion of measuring distances.} in moduli space larger than $\Delta$, the effective theory breaks down which is signalled by a tower of massive states becoming exponentially light \cite{Ooguri:2006in, Klaewer:2016kiy, Blumenhagen:2017cxt}. The second, and more recent, one (S2) is a statement about the incompatibility of exact dS vacua in string theory due to quantum breaking \cite{Ooguri:2016pdq,Danielsson:2018ztv,Sethi:2017phn, Brennan:2017rbf}. We do not judge the viability or validity of these constraints; rather, we focus on their implications for inflationary model-building assuming that they have to be satisfied for a consistent low-energy effective theory.

\section{Example: Quintessential brane inflation}

With the usual definition of the slow-roll parameter, $\epsilon_\V \sim \frac{1}{2} \left(\frac{V'}{V}\right)^2$, we see that (S2) implies
\begin{eqnarray}
  \epsilon_\V \sim c^2/2 \sim \mathcal{O}(1),
\end{eqnarray}
ruling out all known models of slow-roll inflation in which the ``slow-roll'' parameter has to be small. One way to bypass this is to assume that the parameter $c$ does not have to exactly $\mathcal{O}(1)$ and only needs to satisfy $c>0$, i.e. these conjectures are parametric \cite{Dias:2018ngv}. However, we do not take this route here. In fact, we first present an explicit model of inflation as an example where one can have a large potential gradient and yet end up with a small slow-roll parameter due to additional friction terms. We shall go on to generalize our arguments later on. 

Let us consider a model of quintessential inflation with a steep potential and a brane correction term \cite{Sahni:2001qp}. In this scenario, the Friedmann equation is modified because of the high-energy corrections (Randall-Sundrum terms) to the Einstein equations on the brane \cite{Shiromizu:1999wj,Maartens:1999hf} and takes the form 
\begin{equation}
H^2=\frac{\rho}{3 \Mpl^2}\left(1+\frac{\rho}{2\lambda_{\rm B}}\right),
\end{equation}
and the slow-roll parameters are modified to \cite{Maartens:1999hf}
\begin{eqnarray}
&&\epsilon_\H=\epsilon_\V\frac{1+V/\lambda_{\rm B}}{\left(1+V/2\lambda_{\rm B}\right)^2}
\label{eq:epsilon}\\
&&\eta_\H=\eta_\V\left(1+V/2\lambda_{\rm B}\right)^{-1},
\end{eqnarray}
where $\lambda_{\rm B}$ is the brane tension and $V$ is the scalar potential. During the slow-roll phase, $\rho\approx V$ and $\epsilon_\V$ and $\eta_\V$ are the standard slow-roll parameters for the canonical scalar field. In the high-energy limit $V\gg \lambda_{\rm B}$ and $\epsilon_\H,\eta_\H\ll 1$ despite the fact that  $\epsilon_\V,\; \eta_\V$ are large due to the large slope of the potential.

The motivation to consider the potential to be steep, e.g. of an exponential shape
\begin{equation}
 V(\phi)= V_0 \e^{\alpha\phi/\Mpl}
 \label{eq:pot}
\end{equation}
is that at the end of inflation, the scalar field rolls very fast and its energy density decays quickly ($\rho_\phi\sim1/a^6$). So when the universe enters the radiation dominated epoch, the scalar field becomes sub-dominant and does not affect nucleosyntheis. However, during late times, this scalar field can provide acceleration, as a quintessence field, by imposing additional conditions on the potential. For instance, if one considers
\begin{equation}
V=V_0\Bigg[\cosh\(\frac{\gamma \phi}{\Mpl}\)-1\Bigg]^p,~~{\rm where}~~p>0 \, ,
\end{equation}
the potential behaves like an exponential one~(\ref{eq:pot}) with $\alpha\equiv \gamma p$ for large field during the inflationary era, while around the origin $V\sim \phi^{2p}$ such that the average equation of state is $\langle 
w_\phi\rangle=(p-1)/(p+1)$. Setting $p$ small, the equation of state is close to $-1$ during late times. In this scenario, one can connect inflation and late time acceleration with a single scalar field through this quintessential inflation model \cite{Hossain:2014zma}. Since the potential~(\ref{eq:pot}) does not have a minima, the standard reheating mechanism can not be applied. However, an instant preheating mechanism can be used for this kind of non-oscillatory models \cite{Felder:1998vq,Felder:1999pv}. Incidentally, the swampland criteria also favors quintessence models at late times  \cite{Heisenberg:2018yae} while ruling out the standard $\Lambda$CDM scenario\footnote{Although other observational data might be in some tension with quintessence models \cite{Colgain:2018wgk,Dutta:2018vmq,Akrami:2018ylq}.}. However, we shall not get into details of this here and return to our discussion on inflation.

The crucial feature of this model, for our purposes, is that $\epsilon_\H \ll 1$ even when $\epsilon_\V \sim \mathcal{O}(1)$, thus satisfying (S2). Of course, this is not an exclusive feature of this model alone,  which we present only as an example to make our case, but rather of a class of models with additional frictional terms allowing for the potential gradient to be large. However, a general feature of such models, having a canonical scalar field, with fast-roll (under a steep convex potential) is that they generically predict a large tensor-to-scalar ratio, $r$, in contradiction with current data. In this specific scenario, $r\simeq 24\epsilon_\H$ in the high energy limit ($V\gg \lambda_{\rm B}$) \cite{Bento:2008yx}. For potential~(\ref{eq:pot}), $r=24/(N+1)$, where $N$ is the number of e-folds and assuming  $N\sim 60$ we have $r\approx 0.4$ which is ruled out by the Planck 2015 data \cite{Ade:2015lrj}. In fact, this has been argued more generally in \cite{Singlefield1} and shown that single-field models of inflation are severely disfavored once we consider (S2) and observational constraints -- on one hand, if one has a concave potential $V''<0$, then (S2) rules them out and on the other hand, for a convex potential, like in our case, one has observationally ruled out values for $r$. 

This is where we introduce the main idea behind this article in order to avoid these no-go theorems. The observational predictions from inflation depends not only on the specific model and shape of the potential but also on the initial conditions for perturbations. First, we show how an excited initial state for perturbations with a non-Bunch-Davies (NBD) component can resolve the observational tension for the specific model we have discussed thus far. Then, we shall generalize the discussion to show how a NBD initial state is precisely what can help single-field models of inflation in reconciling with the Swampland criteria and observational bounds.
 
\section{NBD initial states and $r$}

The usual assumption of having scalar and tensor perturbation modes in the Bunch-Davies vacuum at the onset of inflation may not be true due to complicated pre-inflationary dynamics \cite{Danielsson:2002kx, Agarwal:2012mq, Kundu:2011sg, Kundu:2013gha}. When thinking of inflationary models as low-energy effective theories, it is natural to assume that inflation started at some finite initial time $\eta_0$, before which one has to take recourse to some UV completion due to quantum gravity \cite{Danielsson:2002qh}. This is especially relevant to the discussion at hand since we are interested in the constraints imposed on inflation due to consistency conditions coming from quantum gravity. In this case, it is entirely possible that deviations from Bunch-Davies results from a previous phase of anisotropic expansion \cite{Dey:2011mj,Dey:2012qp}, a non-attractor solution \cite{Lello:2013mfa}, a cut-off energy scale \cite{Martin:2000xs,Kempf:2000ac}, tunneling from a false vacuum \cite{Sugimura:2013cra}, multi-field dynamics \cite{Shiu:2011qw} or a specific quantum gravity proposal \cite{Agullo:2015aba}. 

An initial generalized state for the scalar ($\zeta$) and tensor ($h$) fluctuations can be usually parametrized as a Bogoliubov transformation of the Bunch-Davies one as
\begin{eqnarray}
	\zeta_{\bf k}(\eta) &=& v_k^{(s)} (\eta) a_{\bf k} + v_k^{(s)\star} (\eta) a_{\bf k}^\dagger,\\
	h_{\bf k}^p(\eta) &=& v_k^{(t)} (\eta) a_{\bf k}^p + v_k^{(t)\star} (\eta) a_{\bf k}^{p\dagger},
\end{eqnarray}
where the mode functions $v^{(s),(t)}$ incorporates the Bogoliubov rotations on both the scalar and tensor Bunch-Davies modes
\begin{eqnarray}
	v^{i}_k(\eta) = \alpha_k^i u_k^i(\eta) + \beta_k^iu^{i\star}_k(\eta)\,,
\end{eqnarray}
where $i$ stands for both scalar (s) and tensor (t) modes. Here, $a_{\bf k}, a_{\bf k}^\dagger$ stand for the usual creation and annihilation operators and $p$ labels graviton polarization. The Bunch-Davies modes are the usual ones
\begin{eqnarray}
	u^{(s)}_k(\eta) &=& \frac{H^2}{\dot{\varphi}} \frac{1}{\sqrt{2 k^3}} (1+ik\eta) e^{-ik\eta},\\
	u^{(t)}_k(\eta) &=& \frac{H^2}{M_p} \frac{1}{\sqrt{ k^3}} (1+ik\eta) e^{-ik\eta}.
\end{eqnarray}
The canonical commutation relation imposes the normalization condition on the Bogoliubov coefficients 
\begin{eqnarray}
	|\alpha_k^i|^2 - |\beta_k^i|^2 = 1\,.
\end{eqnarray}
It is then straightforward to calculate the power spectra of both scalar and tensor perturbations in terms of these excited initial states, and to derive \cite{Hui:2001ce,Ganc:2011dy} (For our notation, see \cite{Brahma:2013rua})
\begin{eqnarray}
r = 16 \epsilon_\H \frac{|\alpha_k^{(t)} + \beta_k^{(t)}|^2}{|\alpha_k^{(s)} + \beta_k^{(s)}|^2} =: 16 \epsilon_\H \gamma\,,
\end{eqnarray}
where the factor $\gamma$ is not a completely free parameter as the NBD component to the initial state of both scalar and tensor modes are constrained by backreaction as well as non-Gaussianity constraints.

A detailed analysis of the constraints appearing on the allowed parameter range of the Bogoliubov coefficients $(\alpha, \beta)$ had been carried out in a series of works (see, for instance \cite{Ashoorioon:2013eia,Flauger:2013hra}). Here, we quote some of their main results relevant for our discussion. As shown in \cite{Ashoorioon:2013eia}, the suppression factor can be easily made to be of order $0.1$ or even lower, albeit being completely consistent with current data. If one assumes a crude exponential model with a very steep cut-off $M$, following \cite{Ashoorioon:2013eia,Holman:2007na}, $\beta^{(s)}_k \propto \exp{(-k^2/M_s^2a^2)}$ \footnote{Another way to model NBD states would be a power law parametrization $|\beta_k|^2 \propto (k/M_s)^{-4-\delta}$.}, this would imply that one gets $\gamma \le 1$ regardless of the details of the initial state for tensor modes. Of course, one can fine-tune $\gamma$ to be much smaller by playing with the relative phase between $\alpha$ and $\beta$, as well as implementing the NBD state through a more sophisticated mechanism. (It is also possible to consider more general states, as in \cite{Agarwal:2012mq}, which do not have to be Bogoliubov rotations of Bunch-Davies, or even Gaussian.) The relative phase for scalar modes  $\Theta^{(s)}$ is defined through  $|\alpha_k^{(s)} + \beta_k^{(s)}|^2 = 1 + 2 |\beta_k^{s}|^2 + 2\sqrt{|\beta_k^{(s)}|^2\left(|\beta_k^{(s)}|^2+1\right)}\, \cos\Theta^{(s)}$, and similarly for tensor modes. Indeed, it has been shown that backreaction constraints on $\beta^{(s)} < 0.1$ do not apply in the $\Theta^{(s)} = \pi$ case \cite{Flauger:2013hra}. This allows us to find parts of the parameter space of NBD states where $\gamma < 0.1$ and we only present this number as an illustrative example.

However, as has been nicely pointed out in \cite{Ashoorioon:2018sqb}, this part of the parameter space of the NBD states has been ruled out by constraints on non-Gaussianity in the scalar $3$-point function. If one chooses the tensor modes to be in the BD vacuum (i.e. the only suppression in $\gamma$ comes from scalar NBD modes), then the non-Gaussianity constraints restricts the upper limit on $c \le 0.2$, which is still too small to be compatible with (S2). However, it is possible to go around this problem by adopting the exact opposite attitude -- keeping the scalar perturbations in the BD vacuum while increasing the NBD component of the tensor modes.  Even considering the bounds on the primordial gravitational wave bispectrum, the allowed part of the NBD parameter space of tensor modes is still large enough to have $c\sim 0.8$ \cite{Ashoorioon:2018sqb}. In fact, albeit not mentioned in \cite{Ashoorioon:2018sqb}, even larger values of $c \sim 0.9$ are also allowed without affecting the backreaction constraints which require that the scale of new physics is sufficiently larger than the Hubble scale of inflation $M_t < H$. Since our aim is to have values of $\gamma < 1$ by increasing the contribution of NBD states for tensor modes, we naturally need to have the phase angle $\Theta^{(t)} > \pi/2$, such that the phase between $\alpha^{(t)}$ and $\beta^{(t)}$ is opposite. This, in turn, is responsible for suppressing the enhancement to the three-point function of tensor modes due to the initial NBD state \cite{Ashoorioon:2018sqb} and thus obeys current bounds on non-Gaussianity.

Our main point is that one can easily consider a NBD initial state for tensor, and even to a smaller degree in the scalar, perturbations which generically lead to a suppression factor in the tensor-to-scalar ratio obtained for inflationary models. The parameter space of these excited states which allows for such a suppression is completely consistent with both theoretical backreaction constraints and observational bounds from primordial non-Gaussianity \cite{Ashoorioon:2013eia, Ashoorioon:2018sqb}. In the specific case of the model described above, the value of $r$, if calculated for one such NBD state, can get suppressed by a factor of $\gamma\le 0.1$ and then $r \sim 0.04$. This makes this model phenomenologically viable once again. Of course, our aim is not to rescue a specific model of inflation but rather to point out that NBD states can be used as a general strategy to reconcile models of inflation with the Swampland conjectures. 

\section{NBD states and Swampland criteria} As pointed out in \cite{Agrawal:2018own,Kehagias:2018uem}, the main problem with (S2) comes to light when one considers the observed values of the $r$. Since $r\le 0.07$ is tightly constrained from experiments \cite{Ade:2015lrj,Ade:2015tva}, the single-field consistency condition \cite{Maldacena:2002vr}, $r =16 \epsilon_\H$, implies that $\epsilon_\H \le 4.4 \times 10^{-3}$. This leads to very small values of $c \le \mathcal{O}(0.1)$. A natural way around this would be to consider models of inflation which violate this consistency relation mentioned above. This inevitably requires going beyond `single-clock' inflation by adding new degrees of freedom in one way or another such that the spectra of the perturbations are not completely specified by the quasi-dS background evolution. For instance, this route has already been proposed in \cite{Kehagias:2018uem} by adding a curvaton field to satisfy (S2). In this work, sticking to single-field models, we present a complementary approach which does not require any additional fields for the duration of inflation -- a NBD initial state for the fluctuations.

As already demonstrated in the previous section, a NBD state implies a modified consistency relation between $r$ and $\epsilon_\H$. Due to the presence of the suppression factor $\gamma$, one can have a larger value of $\epsilon_\H$ which is still consistent with the observed bounds on $r$. For instance, if $\gamma \sim 0.01$, then even for $c \sim 0.9$, one would get $r < 0.07$. Of course, one might ask if it would be possible to get a value of $\gamma$ which is as small as $\mathcal{O}(0.01)$ consistent with observations. Once again, referring to the crude (exponential) model described above, this implies the scale of new physics would be around $M \sim 20 H$\footnote{However, as mentioned earlier, we need to consider most of the NBM component to be in the tensor modes and very little in the scalar ones due to non-Gaussianity constraints.}. We refer the inquisitive reader to \cite{Ashoorioon:2013eia,Ashoorioon:2018sqb} for details of these statements. Thus, it is entirely possible to have a $\mathcal{O}(1)$ number for $c$, as required by (S2), if one introduces NBD states for inflation. At the very least, we have shown that NBD states typically provide a suppression factor $\gamma$ in the expression for $r$ in terms of $\epsilon_\H$, thereby allowing for much larger values for the latter. More generally, we have shown a loophole in the arguments of \cite{Singlefield1} for single-field models. Naturally, our argument is strengthened for models with a non-trivial speed of sound as is immediately clear from the relation between $r$ and $\epsilon_\H$ for such models \cite{Cheung:2007st}.

\section{Small-field excursion conjecture}

So far, we have not said much regarding the first Swampland criterion (S1). It is so because it has been shown in \cite{Agrawal:2018own,Kehagias:2018uem} that there are  a large class of inflationary  models which can be made consistent with (S1), provided the number of $e-$folds is not too large. In fact, as shown in \cite{Agrawal:2018own}, even assuming a scaling behaviour of the form $\epsilon_\H \sim N^{-k}, 1< k <2$, one gets a bound of $\Delta \sim 5$ for plateau potentials. Moreover, once again, the slow-roll parameter might be made small by some higher curvature correction (as in our explicit example) without requiring the potential gradient be very small. This type of model would naturally not be in conflict with (S1) as has known to be the case \cite{Hossain:2014ova}. For the exponential potential~(\ref{eq:pot}), the field excursion during inflation is $\delta\phi/\Mpl=\ln (N+1)/\alpha$ and for $N\sim 60$, we have $\delta\phi\approx 4.11/\alpha$. As already mentioned, in this scenario, the scalar field  energy density remains non-zero even during the post-inflationary dynamics and the scalar field can play an important role in late-time cosmology. For potential~(\ref{eq:pot}), the fractional energy density of the scalar field during radiation dominated era is $\Omega_\phi=4/\alpha^2$ which is highly constrained from big bang nucleosynthesis. For an estimate, if $\Omega_\phi\leq0.01$ then $\alpha\geq20$ implying $\delta\phi/\Mpl\leq 0.2055$, which is explicitly sub-Planckian \cite{Hossain:2014ova}.

Although from an explicit calculation one can obtain that the field range traversed by the inflaton is explicitly sub-Planckian for this model, as mentioned above, it is nevertheless important to ask how the Lyth bound is modified for such models. There are two sources due to which the standard Lyth bound gets modified for our case: Firstly, the presence of NBD initial states leads to a factor of $\gamma$ correction whereas presence of additional `brane-correction' terms means that the $\epsilon_\H \neq \epsilon_\V$. Taking both of these into account, the new Lyth bound is given by \cite{Hossain:2014ova}
\begin{equation}
 \delta\phi\gtrsim \sqrt{\frac{r}{24\gamma\(N+1\)}}\frac{N}{\alpha}\Mpl \, .
 \label{eq:lyth_bound_RS}
\end{equation}
Due to the NBD states, the $\gamma$ in the denominator tries to increase the range of the field excursion, $\gamma$ being less than $1$ as required by (S2), but the factor of $\alpha$ shows the presence of the specific (steep) potential appearing in this expression due to the high-energy corrections. (In the canonical case, there is no explicit presence of any parameter associated with the potential since $\epsilon_\H = \epsilon_\V$ in that case.) Also there is an additional factor of $1/(3N+3)$ under the radical sign due to the same corrections as compared to the canonical expression for the Lyth bound. Therefore, we can easily satisfy the (modified) Lyth bound in our model in spite of NBD initial states.

The fact that (S1) typically requires inflationary models to not have too much inflation, i.e. not much beyond what is required from experiments ($\sim 60$ $e$-folds) also bodes well for our hypothesis of NBD initial states. In general, if there are many $e$-folds of inflation, one can lose any detectable traces of such deviations from BD in the initial state of perturbations\footnote{It would be like taking the limit $\eta_0\rightarrow 0$.}. However, if inflation indeed started at some finite time, as required by (S1), it makes the likelihood of having a NBD component for the initial state a lot more likely due to complicated pre-inflationary dynamics, such as a quantum gravity era with additional terms modifying the effective action above some scale.

\section{Conclusion}

In this work, we focussed on how to make models of single-field inflation compatible with the swampland criteria. Firstly, there are known models where one considers higher curvature corrections to Einstein's equations, leading to additional friction terms in the Friedmann and Klein-Gordon equations. This results in the Hubble slow-roll parameter ($\epsilon_\H$) to be \text{not} approximately equal to $\epsilon_\V$, and they are related via additional correction terms as in (\ref{eq:epsilon}). This gives us a natural way to have a $\mathcal{O}(1)\; \epsilon_\V$, as required by the swampland criterion (S2), even though $\epsilon_\H$ can be small. The specific model we consider in this article has the additional advantage of integrating inflation with late-time quintessence with the same scalar field. The string swampland requires that we explain our current universe via quintessence \cite{Agrawal:2018own,Heisenberg:2018yae} rather than by an exact dS space and therefore, we shall explore the viability of this model in this regard in future work. (However, note that there have been some interesting contradictions noted against such quintessence models obeying (S2) when the Higgs field is considered \cite{Denef:2018etk}.)

On the flip side, these models are typically ruled-out by observational bounds on $r$. This is where we input our crucial ingredient of having a NBD initial state for cosmological perturbations\footnote{The issue of a NBD vacuum is also at the heart of the effect of quantum loop corrections to the swampland conjectures \cite{Danielsson:2018qpa}.}. This results in suppressing the value of $r$ by a considerable amount to make these models compatible with data. Finally, this gives us a fresh route towards making single-field inflationary models compatible with the swampland. Since the most restrictive bounds on these models come from the consistency relation between $r$ and $\epsilon_\H$, one can easily violate it by using NBD states. These, of course, cannot be considered as single-clock inflationary models anymore. However, as explained earlier, it is quite natural to expect some general NBD states if there was some pre-inflationary quantum gravity regime with inflation beginning at some finite time. There are several falsifiable observational signatures of NBD states from future data on the non-Gaussian statistics of primordial fluctuations. Thus, in this work, we present a path towards incorporating models of inflation with string theory as well as explicitly laying down the testable phenomenological characteristics of our assumptions.

\section*{Acknowledgments} We thank Eoin Ó. Colgáin and Dong-han Yeom for useful comments on an earlier version of this draft. We are also grateful to Arthur Hebecker for pointing us towards the Higgs problem for the swampland and for stimulating discussions on the issue and to the anonymous referee for clarifications on the relevant non-Gaussianity constraints for excited scalar perturbations. 

\noindent This research was supported in part by the Ministry of Science, ICT \& Future Planning, Gyeongsangbuk-do and Pohang City  and the National Research Foundation of Korea (Grant No.: 2018R1D1A1B07049126).

\end{document}